# NEW PRINCIPLES OF RADIATION DAMAGE AND RECOVERY BASED ON THE RADIATION INDUCED EMISSION OF SCHOTTKY DEFECTS

Vladimir I. Dubinko
*NSC Kharkov Institute of Physics and Technology, Kharkov 310108, Ukraine*

## Abstract

In the present paper, a mechanism of radiation-induced emission of Schottky defects from extended defects is proposed, which acts in the opposite direction compared with Frenkel pair production, and it results in the *radiation-induced recovery* processes. The vacancy emission from a void surface results in a *shrinkage* of the void which is analogous to the thermal void shrinkage at high temperatures, but it is expected to operate at low and medium temperatures and high sink densities, when the radiation-induced vacancy emission from sinks becomes comparable with the absorption of Frenkel defects produced in the bulk. As a result, the void growth becomes *negative* below a threshold temperature (or above a threshold flux), and *saturates* after a threshold irradiation dose, at a level, which depends on the type of material. These effects have been observed experimentally for a long time but could not be explained by the conventional theory. One of the technologically important consequences of the new concept is a mechanism of irradiation creep, based on the radiation and stress induced preference emission (RSIPE), which is essentially temperature independent in contrast to the well-known SIPA or elastodiffusion mechanisms that can yield a significant irradiation creep only when recombination is negligible. The efficiency of the radiation-induced Schottky defect emission is different for different types of materials, which is shown to be one of the major causes of their different responses to irradiation.

# 1. Introduction

In the conventional theory of radiation damage formulated more than 30 years ago in terms of the mean-field chemical rate equations [1], it is assumed that under irradiation, Frenkel pairs of vacancies and self-interstitial atoms (SIAs) are created in the bulk and annihilate at extended defects (ED), which are thus considered primarily as the *sinks* for freely migrating point defects and their small mobile clusters [2, 3]. It is generally assumed that the local concentrations of point defects at the ED surfaces are determined by *thermodynamic equilibrium* conditions. However, as it will be shown in the present work, the surfaces of ED can be a source for the production of *radiation-induced* Schottky defects that does not exist in the bulk. A Schottky defect is a single vacancy or SIA, or a *small defect cluster*, which can be emitted from ED surface and which does not require a counterpart of the opposite sign in contrast to the bulk production of primary defects, in which the total numbers of vacancies and SIAs *must be equal*. By this mechanism, which had not been previously realized, ED can act not only as *sinks* for Frenkel defects but also as the *radiation-induced sources* of Schottky defects, which results in a new driving force for the microstructural evolution; it acts in the opposite direction compared with Frenkel pair production, and it results in the *radiation-induced recovery* processes.

In the present paper, one of the possible mechanisms of the radiation-induced emission of Schottky defects from ED is proposed, which is based on the interaction of ED with *focusing collisions* (focusons) similar to the mechanism of Frenkel pair production at dislocations and stacking faults proposed by Leibfried in early sixties [4]. But it contrast to the Leibfried mechanism, the collision events in the vicinity of ED are expected (due to the lower energy barrier involved) to result in a *biased emission of vacancies* at a rate that depends on the *type* of extended defect. And in contrast to thermal vacancy emission, the radiation-induced one depends on the *dose rate* and *microstructure.*

Physical consequences of the proposed mechanism are numerous, and only some of the most technologically important examples are considered in the present paper, which is organized as follows.

In the section 2, the *focuson mechanism* of radiation Schottky defect emission (RSDE) is developed and compared with Frenkel defect production in the bulk and at ED.

In the section 3, the RSDE mechanism is incorporated in the rate theory along with the thermal Schottky defect emission (ThSDE).

In the section 4, we consider the radiation-induced evolution of dislocation density, which is shown to saturate at a level determined by the RSDE and ThSDE mechanisms *regardless of the initial state*, in agreement with experimental observations.

In the section 5, a mechanism of *irradiation creep* based on the radiation and stress induced preference *emission* (RSIPE) is proposed and compared with conventional mechanisms based on the stress induced preference *absorption.*

In the section 6, the temperature, flux and dose dependence of void swelling is considered. It is shown that the void growth becomes *negative* below a threshold temperature (or above a threshold flux), and *saturates* after a threshold irradiation dose at a level determined by the RSDE mechanism, which strongly depends on the type of material, in agreement with experimental observations. A focuson mechanism of the void ordering is shortly discussed.

Effects due to *impurities*, alloying additions and *sub-threshold* electron irradiation effects are discussed and summarized in the section 7 as well as some general aspects of the proposed theory.

## 2. Focuson mechanism of Schottky defect emission

It has been pointed out by Silsbee [5] and Leibfried [4] that focusing collisions may play an important role in radiation damage. These collisions *transfer energy* along close packed directions of the lattice, but there is *no interstitial transport* by a focusing collision, which enlarges their range considerably as compared to the mean free path of a primary knock-on atom (PKA) in a *crowdion* configuration. A focusing collision has many similarities with a particle because the collision energy is concentrated in a region of about one atom spacing, and has therefore been called shortly a *focuson* [4].

The energy range in which focusons can occur has an upper limit $E_F$, which is estimated to be about 60 eV for Cu [4]. A focuson loses its energy continuously, which determines its *propagation range* as follows. If the energy loss per hit is $\varepsilon_F \approx 10^{-2}$ [4] then a focuson with initial energy $E$, has the energy $E(x)$ after propagating a distance $x$ given by

$$E(x) = E \exp(-x/l_F), \quad l_F = b/\varepsilon_F \tag{1}$$

where $l_F \approx 100b$ is the characteristic propagation range of a focuson in a *perfect crystal*, and $b$ is the atom spacing. Accordingly, a distance at which a focuson of initial energy $E$ has the energy $E_f$ in a *perfect crystal* is given by

$$l_0(E, E_f) = l_F \ln(E/E_f) \tag{2}$$

In an ideal lattice a focuson travels along a close packed direction through a distance corresponding to its initial energy and *does not produce defects* along its path (Fig. 1a). The initial energy is completely transformed into lattice vibrations or heat. However, if the focuson has to cross a region of lattice disorder, a defect may be produced in its proximity. So far only one possible scenario has been considered by Lebfried [4], namely, a *Frenkel pair production* due to a focuson crossing a stacking fault region, which requires fairly high threshold energy $E_d$. By this mechanism, the enhancement of the radiation damage rate by cold work was expected to occur through the enhanced production of Frenkel pairs at dislocations, which can be significant only at extremely high dislocation densities of about $10^{13} cm^{-2}$ [4].

We shall consider another mechanism of defect production, which can operate only at the ED surfaces (such as the grain or the void surfaces and edge dislocations), which can act both as *sinks* and a *sources* of *Schottky defects* that do not require the counterparts of the opposite sign in contrast to the production of Frenkel pairs (Fig. 1b).

The energy given by eq. (1) can be transferred to the atoms at ED surfaces, which a focuson encounters after traveling a distance $x$, and which act as the focuson sinks. As a result of these collisions, Schottky defects (s-defects), such as vacancies, self-interstitial atoms (SIAs) and their small clusters can be produced in a close proximity of ED, in quantities depending on the transferred energy and the ED type. Accordingly, a crystal with ED can be considered as an effective *loss medium* for focusons and an effective *production medium* for s-defects.

The number of focusons produced by a PKA with energy, $E_p$, in the energy range $(E, dE)$ is $n_{F0}(E_p, E) = z(E_p, E)dE$, where $z(E_p, E)$ is given approximately as [4]

$$z(E_p, E) = \frac{2E_p}{E_F^2} \ln \frac{E_F}{E}, \qquad E \leq E_F. \tag{3}$$

The number of focusons having initial energy $E$, which would propagate a distance x in the effective medium and still have the energy $E_f$ can be evaluated from the following equation

$$dn_F(E_p, E, E_f, x) = -\frac{n_F(E_p, E, E_f, x)dx}{l_{ef}(E, E_f, l_{ED})} \tag{4}$$

where $l_{ef}(E, E_f, l_{ED})$ is the mean free path of a focuson propagating one-dimensionally in a crystal containing ED, such as *cavities* (three-dimensional sinks), *dislocations* (linear sinks) and *grain boundaries* (planar sinks), which is determined by the following relations

$$l_{ef}(E, E_f, l_{ED})^{-1} = l_0(E, E_f)^{-1} + l_{ED}^{-1} \tag{5}$$

$$l_{ED}^{-1} = l_C^{-1} + l_D^{-1} + l_G^{-1}, \quad l_C = (\pi R_C^2 N_C)^{-1}, \quad l_D = z_c (2r_{FD}\rho_D(z_c - 2))^{-1} \tag{6}$$

where $l_G$ is the grain size, $R_C$ and $N_C$ are the cavity radius and number density, respectively, $\rho_D$ is the dislocation density, $r_{FD}$ is the focuson – dislocation cross-section radius, and $z_c$ is the host lattice coordination number, which takes into account that a focuson cannot cross dislocations lying along its propagation direction.

From eq. (4) one obtains

$$n_F(E_p, E, E_f, l_{ED}, x) = n_{F0}(E_p, E) \exp\left(-\frac{x}{l_{ef}(E, E_f, l_{ED})}\right), \quad (7)$$

which can be used to evaluate the number of focusons with initial energy $E$ that would *transfer the energy* exceeding $E_f$ to ED of a particular type $S = C, D, G$ as follows

$$n_{FS}(E_p, E, E_f, l_{ED}) = \frac{n_{F0}(E_p, E)}{l_S} \int_0^\infty \exp\left(-\frac{x}{l_{ef}(E, E_f, l_{ED})}\right) dx = n_{F0}(E_p, E) \frac{l_{ef}(E, E_f, l_{ED})}{l_S} \quad (8)$$

Accordingly, the number of focusons absorbed by all ED is given by a sum

$$n_{FED}(E_p, E, E_f, l_{ED}) = n_{F0}(E_p, E) \sum_S \frac{l_{ef}(E, E_f, l_{ED})}{l_S} = n_{F0}(E_p, E) \frac{l_0(E, E_f)}{l_0(E, E_f) + l_{ED}} \quad (9)$$

It can be seen that at high ED densities, $l_0(E, E_f) \gg l_{ED}$, it is close to the total number of focusons produced by a PKA, whereas at low ED densities it is proportional to the ratio $l_0(E, E_f)/l_{ED} \ll 1$.

Now, the type and number of s-defects formed by one focuson encountering an extended defect of type *S* is generally a function of transferred energy and ED type. In the simplest case of *point s-defect emission* that we consider, it is equal to one if the transferred energy is larger than $E_{sS}$ and zero otherwise, where $E_{sS}$ is the minimum energy required to form one point s-defect of type $s = i, v$ at ED of type $S$, and $i, v$ correspond to SIAs and vacancies, respectively. In this case, the total number of s-defects emitted by one PKA from ED of *S*-type, $N_{sS}$, is given simply by the integration of eq. (8) with $E_f = E_{sS}$ over the focuson energy spectrum, which yields

$$N_{sS}(E_d, E_F, l_F, l_S, E_{SF}, l_{FD}) = \frac{2E_d}{E_F} \frac{l_F}{l_S} \Phi_F(E_{SF}, l_{FD}), \quad (10)$$

$$\Phi_F(E_{SF}, l_{FD}) = \int_1^{E_{fF}} \frac{\ln(x/E_{SF})\ln x \, dx}{1 + x l_{FD} \ln(x/E_{SF})}, \quad E_{SF} = E_{sS}/E_F, \quad l_{FD} = l_F/l_{ED} \quad (11)$$

To compare the emission rate of s-defects, $K_{sS}$, with the production of Frenkel pairs (FP), $K_{FP}$ we shall use the simplest estimate for the number of displacements produced by a PKA as [4]

$$N_K(E_p, E_d) = \frac{E_p}{2E_d} = \frac{K}{flux}, \quad (12)$$

where $K$ is the displacement rate measured in $dpa/s$ and $flux$ is the flux of irradiating particles. The production rate of freely migrating FP, $K_{FP}$, was shown to be only a small fraction of the displacement rate under cascade damage. Literature values lie between 1 and 8% [6]. We assume a fraction, $k_{ef}$ of 5% together with the simplification that the numbers of freely migrating SIAs and vacancies are equal, i.e. we will not take into account the *production bias* effects in this paper. In this way we have

$$K_{FP}(K, k_{ef}) = K \times k_{ef}, \quad (12)$$

$$K_{sS}(K, E_d, E_F, l_F, l_S, E_{SF}, l_{FD}) = K \times \frac{4E_d}{E_F} \frac{l_F}{l_S} \Phi_F(E_{SF}, l_{FD}), \quad (13)$$

Figure 2a shows a comparison between production rates of vacancies and FP at dislocations per unite volume. It can be seen that due to the in-cascade recombination they become comparable with the FP production rate in the bulk at much lower dislocation densities, than with the total displacement rate, but in fact, this is not a relevant comparison because s-defects are produced in a *highly localized* way in a close proximity of ED surfaces, and so a large part of them can be subsequently *absorbed* by the ED from which they have been *emitted*. It means that in order to understand the effects caused by RSDE we should incorporate it in the rate theory in a regular way together with the ThSDE, which will be done in the following section.

## 3. Rate theory with account of RSDE

In the conventional rate theory, the local concentrations at the ED surfaces are usually assumed to be determined by *thermodynamic equilibrium* conditions in the diffusion-limited case when there are no special barriers hampering the transfer of point defects to sinks. We will show that even in the diffusion-limited case, this assumption is not valid in the technologically important range of irradiation temperatures and dose rates.

### *3.1. Local equilibrium concentrations at extended defects under irradiation*

In the rate theory, the sink strengths are obtained by solving diffusion equations for the "sinks" embedded in an effective medium with boundary conditions of the general type at the interior surface defined by a particular ED:

$$j_{sS} = v_s \left[ c_s(S) - c_{sS}^{eq} \right], \tag{14}$$

where $j_{sS}$ is the component of the point defect flux along the normal at the ED-matrix interface, directed internally towards the ED, $v_s$ is the transfer velocity from the matrix to ED or vice versa, $c_s(S)$ is the local PD concentration near the ED surface, and $c_{sS}^{eq}$ is the local *equilibrium* concentration so that the flux is zero when $c_s(S) = c_{sS}^{eq}$. Without irradiation, one has $c_{sS}^{eq} = c_{sS}^{th}$, which can be obtained from thermodynamics by minimizing the free energy of the system containing identical ED. The product $v_s c_{sS}^{th}$ is the *thermal emission rate* from a particular ED. Under irradiation, $c_{sS}^{eq}$ is generally given by the sum of $c_{sS}^{th}$ and $c_{sS}^{irr}$ due to the radiation-induced emission of s-defects, where $c_{sS}^{irr}$ has a purely kinetic origin and cannot be obtained from thermodynamic considerations. However, one can evaluate the *radiation-induced emission rate*, $v_s c_{sS}^{irr}$, which is given by the ratio of the emission rate from all ED of S-type, $K_{sS}$ (eq. (13)), to the total ED surface per unite volume. In the case of dislocations, one obtains in this way

$$v_s c_{sD}^{irr} = \frac{K_{sD}}{2\pi r_{DF} \rho_D} = \frac{4(z_c - 2)}{\pi z_c} \times \frac{E_d}{E_F} l_F \Phi_F (E_{DF}, l_{FD}), \quad E_{DF} = E_{sD}/E_F \tag{15}$$

The maximum transfer rate (when there are no special barriers at ED surfaces) is given approximately by $v_s = D_s/b$, where $D_s$ is the point defect diffusivity, which is also a sum of the thermal and radiation-induced constituents:

$$D_s = D_s^{th}(E_{sM}, T) + D_s^{irr}(K, E_d, E_F, l_F, E_{MF}, l_{FD}), \tag{16}$$

$$D_s^{irr}(K, E_d, E_F, l_F, E_{MF}, l_{FD}) = K_{sM}(K, E_d, E_F, l_F, E_{MF}, l_{FD}) \times b^2 \times f_c^m, \tag{17}$$

$$K_{sm}(K, E_d, E_F, l_F, E_{mF}, l_{FD}) = K \times \frac{E_d}{E_F} \frac{l_F}{b} \Phi_F(E_{mF}, l_{FD}), \quad E_{mF} = E_{sm}/E_F \tag{18}$$

where $E_{sm}$ is the migration energy, and $f_c^m$ the migration kinetic factor presented in the table of material parameters.

Substitution of $v_s$ into eq. (15) yields an expression for the *radiation-induced local equilibrium concentration* of s-defects at dislocations

$$c_{sD}^{irr} = K_{sD}^{loc}(K, E_d, E_F, l_F, E_{DF}, l_{FD}) \times \frac{b^2}{D_s} \times f_c^D, \quad f_c^D = \frac{4(z_c - 2)}{\pi z_c} \tag{19}$$

$$K_{sD}^{loc}(K, E_d, E_F, l_F, E_{DF}, l_{FD}) = K \times \frac{E_d}{E_F} \frac{l_F}{b} \Phi_F(E_{DF}, l_{FD}), \tag{20}$$

where $f_c^D$ is the dislocation crystallographic factor, and $K_{sD}^{loc}$ is the *local emission rate* of s-defects, which is compared with the bulk production rate of FP in Fig. 2b. It can be seen that the local emission of vacancies exceeds the bulk production of FP by several orders of magnitude, and in contrast to the total production of vacancies (Fig. 2a) it decreases with increasing dislocation density due to decreasing of the focuson mean free path in a crystal with dislocations. It is useful to remember that the thermal equilibrium can be realized only in a system with *identical* "sinks" without irradiation. In the same way, if thermal emission and the bulk production of FP were *forbidden* ($c_{sD}^{th} \to 0$, $E_d \to \infty$), the mean PD concentrations in a crystal with identical dislocations *under irradiation* would coincide with $c_{sD}^{irr}$, and hence there would be *no flux* across their surfaces, which means that $c_{sD}^{irr}$ is an exact *kinetic analogue* of the thermodynamic equilibrium concentration, which operates in the open, i.e. "non-equilibrium" radiation environment.

The local concentrations at grain boundaries and cavities can be obtained in a similar way, and they have the same structure as $c_{sD}^{irr}$:

$$c_{sS}^{irr} = K_{sS}^{loc}(K, E_d, E_F, l_F, E_{SF}, l_{FD}) \times \frac{b^2}{D_s} \times f_c^S, \qquad f_c^S = 1, \quad S = G, C \tag{21}$$

Comparison of thermal and radiation-induced equilibrium PD concentrations at dislocations for a typical neutron flux (Fig. 3) shows that the latter dominates completely at temperatures below 0.5 $T_m$, where all significant radiation effects are observed. At low temperature end, the thermal equilibrium concentration is practically zero, whereas the radiation-induced equilibrium concentration may exceed the melting point level! For vacancies (Fig. 3a) it saturates when the radiation-induced diffusion dominates over the thermal one.

### 3.2. Radiation-induced emission bias

The bias is a central concept in the theory of radiation damage. The presence of the bias differentials between at least two kinds of sinks causes an imbalance of SIA and vacancy fluxes to each of the sinks, resulting in microstructural evolution and macroscopic deformation.

Until now two general bias types have been considered, namely, *absorption* and *production* biases. The absorption bias may be due to elastic interaction difference between dislocations [7, 8] and voids [9] with SIAs and vacancies, or due to the diffusion anisotropy difference, either intrinsic [10], or stress-induced [11-13]. The production bias is a consequence of different efficiency of production and stability of small vacancy and SIA clusters [2] and dislocation loops [3] produced in displacement cascades.

We shall consider a new type of bias, namely, the *emission bias*, which is caused by the difference in emission rates of vacancies and SIAs by different ED. In fact, the emission bias can operate even without irradiation and result in recovery of microstructure under *thermal annealing*, which is due to the difference in thermal emission rates of vacancies from grain boundaries, voids and dislocations. This difference is caused by different vacancy formation energies at these ED. Emission of SIAs can be neglected in metals due to higher formation energy of SIAs as compared to vacancies.

In a similar way, the radiation-induced emission of vacancies is more efficient than that of SIAs, and it is more efficient for voids and grain boundaries than for dislocations, as shown in Fig. 4a. It should be noted that the vacancy emission by a focuson requires the energy, which may be *different* from that of "thermal" vacancy, and which determination is a promising area of further research. In the present work we have assumed that the "radiation" vacancy formation energies at void and grain boundary surfaces coincide with thermal values, and they are larger for dislocations by a value determined by the stacking fault energy (SFE) and kinetic coefficient listed in the table 1.

SIAs are not expected to be produced at void and grain boundary surfaces, and according to MD simulations [14], "radiation" SIA formation energy at dislocations may even exceed the FP formation energy in the bulk, $E_d$. So like in the thermal case, SIA emission can be neglected, for the sake of simplicity.

Now the ED emission bias can be defined as a relative difference between the equilibrium vacancy concentrations at the ED and a free surface, $c_{v0}^{eq}$:

$$B_{eS} = \frac{c_{vS}^{eq} - c_{v0}^{eq}}{c_{v0}^{eq}}, \quad c_{vS}^{eq} = c_{vS}^{th}(T) + c_{vS}^{irr}(K,T), \quad S = D, C, G \tag{22}$$

For a free grain boundary we assume $c_{vG}^{eq} = c_{v0}^{eq}$, and hence $B_{eG} = 0$. So the emission bias is zero for a free grain boundary, it is positive for voids and depends on their radii, and it is negative for dislocations and depends on their SFE, as shown in Fig. 4b. A void and a vacancy loop shrinks, whereas a SIA loop grows if its emission bias is larger than the mean emission bias of the system, $B_{eM}$, given by

$$B_{eM} = \frac{c_{vM}^{eq} - c_{v0}^{eq}}{c_{v0}^{eq}}, \quad c_{vM}^{eq} = \sum_{S} c_{vS}^{eq} k_{vS}^2 \Big/ k_v^2, \quad k_v^2 = \sum_{S} k_{vS}^2, \quad S = D, C, G, \quad (23)$$

where $c_{vM}^{eq}$ is the mean equilibrium vacancy concentration, and $k_S^2$ is the sink strength of S-type ED defined in the rate theory, which includes now the RSDE mechanism. Below we shall consider some physical consequences of the new theory.

## 4. Radiation-induced evolution of dislocation density

It is well known that the starting dislocation density is not maintained during irradiation and it evolves toward a saturation level that is independent of the starting state [15] but is mainly determined by the material type and, to some extent, by irradiation conditions. For example, in model Fe-Cr-Ni alloys and various 300 series stainless steels, characterized by relatively low stacking fault energy, the saturation density of network dislocations has been shown to be $6 \pm 3 \times 10^{10} cm^{-2}$, "relatively independent of starting state, temperature, dose rate, He/dpa ratio and most other important parameters [15]". This saturation process involves an *order of magnitude reduction* in the dislocation density of typical cold-worked steels and a *comparable or larger increase* in the density of annealed steels as shown in Fig. 5a. Up to date there was no theoretical explanation of this universal behaviour to our knowledge. And indeed, if the dislocation climb velocity is positive under irradiation (as it follows from the conventional theory) then the effect of irradiation should be similar to that of cold-work and result in a comparable or larger dislocation density, which is contrary to the experimental facts.

This contradiction is resolved very easily by taking into account the emission bias effect, which provides a *negative* constituent to the radiation-induced dislocation climb velocity, $V_D$, which can be *comparable or larger* than a positive constituent determined by the absorption bias. Neglecting SIA emission for simplicity, one obtains from the modified rate theory

$$V_D = b^{-1}Z_{iD}D_v\left(c_{vM} - c_{vM}^{eq}\right)(B_{aD} - B_{aM}) - b^{-1}Z_{vD}D_v^{rsd}(B_{eM} - B_{eD}), \tag{24}$$

$$B_{aD} = \frac{Z_{iD} - Z_{vD}}{Z_{iD}}, \quad B_{aM} = \frac{k_i^2 - k_v^2}{k_i^2}, \quad D_v^{rsd} = D_v c_{v0}^{eq}, \tag{25}$$

where $c_{vM}$ is the mean vacancy concentration determined by the rate equations in a usual way [9], $B_{aD}$ and $B_{aM}$ are the dislocation and the mean microstructure *absorption* biases, respectively, $Z_{sD}$ are the dislocation capture efficiencies for PD, which are essentially determined by the relaxation volumes associated with SIAs and vacancies, $\Omega_s$, and given approximately by [9]

$$Z_{sD} = 2\pi \Bigg/ \ln\left(\frac{2}{L_s\sqrt{k_s^2}}\right), \quad L_s = \frac{\mu b(1+\nu)}{3\pi kT(1-\nu)}|\Omega_s|, \tag{26}$$

where $b$ is the host lattice spacing, $\mu$ is the shear modulus of the matrix, $\nu$ is the Poisson ratio, and $kT$ has its usual meaning. $D_v^{rsd}$ is *the radiation-induced* vacancy self-diffusion, which is a function of both temperature and *dose rate*.

Now the evolution of dislocation density with time can be described by a simple rate equation:

$$\frac{d\rho_D}{dt} = \frac{\rho_D}{\tau_D}, \quad \tau_D = \frac{l_{DD}}{V_D}, \quad l_{DD} = \frac{1}{2r_{DD}\rho_D} \tag{27}$$

where $l_{DD}$ is the mean free path of a climbing dislocation before the annihilation with another dislocation having the opposite Burgers vector, and $2r_{DD}$ is the maximum annihilation distance determined by the strength of the dislocation elastic attraction and the threshold glide stress, $\sigma_P$:

$$r_{DD} = \frac{\mu b}{2\pi(1-\nu)\sigma_P} \tag{28}$$

Let us consider dislocation evolution prior to void nucleation in a crystal containing grain boundaries (GB) as the only other alternative ED. The GB sink strength depends on the grain size and the total sink strength of the microstructure within the grain. We consider the case of planar sinks, which are often used to simulate grain boundaries. A pair of parallel planes separated by a distance $2l_G$ has the following sink strength and absorption bias [10]

$$k_G^2 = \frac{k_D}{l_G}\left(\coth[k_D l_G] - \frac{1}{k_D l_G}\right)^{-1} = \begin{cases} k_D/l_G, & k_D l_G \gg 1 \text{ (coarse grains)} \\ 3/l_G^2, & k_D l_G \ll 1 \text{ (fine grains)} \end{cases}, \qquad (29)$$

In the case of coarse grains, neglecting the bulk recombination of PD one gets from eqs. (24) – (27) a very simple equation for the evolution of dislocation density with time,

$$\frac{d(\rho_D/\rho_D^{sat})}{dt} \approx \frac{(\rho_D/\rho_D^{sat})^{1/2}}{\tau_{sat}}\left(1 - \frac{\rho_D}{\rho_D^{sat}}\right), \qquad (30)$$

$$\rho_D^{sat} = \frac{K_{FP}\left(1-\sqrt{Z_{vD}/Z_{iD}}\right)}{D_v^{rsd}Z_{vD}|B_{eD}|}, \quad \tau_{sat} = \frac{bl_G}{2r_{DD}}\left(\frac{Z_{iD}}{K_{FP}D_v^{rsd}Z_{vD}|B_{eD}|\left(\sqrt{Z_{vD}/Z_{iD}} - Z_{vD}/Z_{iD}\right)}\right)^{\frac{1}{2}}, \qquad (31)$$

which results in $\rho_d(t)$ evolution toward a saturation, $\rho_D^{sat}$, *regardless of the initial state*, in agreement with experimental observations, as shown in Fig. 5b. The saturation level (Fig. 6a) is essentially determined by the RSDE below 0.5 $T_m$, where it is practically independent of starting state, temperature, dose rate and most other parameters considered to be "important" in the conventional theory. It decreases with increasing stacking fault energy, which decreases the dislocation emission bias. The saturation dose decreases with decreasing the ratio of the grain size to the annihilation range (Fig. 6b), which explains a more rapid convergence of the dislocation density to $\rho_D^{sat}$ from the cold work as compared to annealed state (Fig. 5).

## 5. Irradiation creep

Irradiation creep is one of the most outstanding puzzles in the theory of radiation damage, because in spite of its technological importance and the consequently large number of attempts to describe it, some of the general trends are still not well understood. It is known that under typical neutron fluxes, irradiation creep dominates over thermal creep below about 0.5 $T_m$, it is closely related to the void swelling between 0.5 and 0.3 $T_m$, and it persists well *below* the swelling temperature range. The experimental data on the swelling- creep relationship [17] indicate that the creep rate, $\dot{\varepsilon}$ per unite stress and dpa may be described by the creep *compliance*, $\dot{\varepsilon}_0$, and another contribution proportional to the swelling rate, $\dot{S}$ [17]

$$\dot{\varepsilon} = \dot{\varepsilon}_0 + s_0 \dot{S} \tag{32}$$

where $s_0 \approx 10^{-2} MPa^{-1}$ and $\dot{\varepsilon}_0 \approx 10^{-6} MPa^{-1} dpa^{-1}$ for a surprisingly large range of austenitic steels, and it shows a rather weak or *no dependence on irradiation temperature*, which can not be understood within the framework of existing creep models. Up to date, there are two main models, namely SIPA [16] and elastodiffusion [12], which are based on the *absorption bias* of dislocations [16] or other ED [12] differently oriented with respect to the external stress. Consequently, these models can yield a significant irradiation creep only when recombination is negligible, and so they can be fitted to provide $s_0$ but not $\dot{\varepsilon}_0$.

We consider another mechanism of irradiation creep based on the radiation and stress induced preference *emission* (RSIPE), which originates from the well known dependence of the vacancy formation energy on the orientation of the dislocations or GB with respect to the external stress field (Table 1). Accordingly, the creep *compliance* is determined by the sum of the dislocation-induced and GB – induced constituents, which are given, respectively by

$$\dot{\varepsilon}_{0D} = \rho_D D_v^{rsd} \left( B_{eD}^{\perp}(\sigma) - B_{eD}^{II} \right) \tag{33}$$

$$\dot{\varepsilon}_{0G} = l_G^{-2} D_v^{rsd} \left( B_{eG}^{\perp}(\sigma) - B_{eG}^{II} \right) \tag{34}$$

where the upper scripts $\perp, II$ correspond to the perpendicular and parallel oriented ED, respectively. Temperature and grain size dependences of the creep rates predicted by the present theory are shown in Fig. 7.

## 6. Void swelling

The temperature dependence of the void swelling rate predicted by the conventional theory is shown in Fig. 7a. It has a well known bell shape and shifts upward along the temperature axis with increasing dose rate; it becomes negative above a threshold temperature (thermal annealing), and it decreases steadily with decreasing temperature (or increasing dose rate) due to the bulk recombination effect, but it does not become negative at the low temperature end.

However, some experiments indicate that voids can shrink with decreasing temperature (or increasing dose rate) after they have been formed under more favorable conditions [18-21]. Steel and Potter [18] reported that voids formed during $Ni^+$ ion bombardment of Ni at 650 C shrink very rapidly when subjected to further bombardment at temperatures between 550 C and 25 C, the rate at which these voids disappear being independent of temperature in this range. Between 650 C and 700 C, the void growth proceeds to fluences near $3 \times 10^{21} ions/m^2$ and is followed by shrinkage.

The authors have attempted to explain the observations using the rate theory modified to include the interstitials injected by the ion beam and sputtering of the surface by the ion beam. However, the former effect is negligible due to a very low production bias introduced by injected SIAs (about 0.1%) as shown in Fig. 8a. The dose required for shrinkage of a void as large as 20 nm at a *maximum rate* induced by SIA injection is about 2000 dpa, which exceeds the maximum dose in this experiment by an order of magnitude.

Figure 8b shows the results of calculations, which neglect the production bias but take into account the increasing dose rate with decreasing distance to the surface as it was reported in [18]. Accordingly, the emission bias increases, which results in the void shrinkage rates well capable to produce the observed effects.

A more general trend of the void swelling is its saturation with increasing irradiation dose, which is a common feature for materials producing relatively high void number densities, such as bcc Mo, Nb and W [20-22]. Among fcc metals, a similar trend was observed for Ni and Al, which have relatively large SFE. There is also evidence (for HFIR irradiation) for the saturation of swelling in Type 316 stainless steel, but at much higher doses and damage levels [20]. Conventional rate theory has no plausible explanations of these phenomena.

Taking into account the RSDE mechanism, the swelling evolution in time can be described in much the same manner as the evolution of dislocation density described in the section 4, which results in the following equation for the void swelling, $S_V$.

$$\frac{d(S_V/S_V^{sat})}{dt} \approx \frac{1}{\tau_{sat}}\left((S_V/S_V^{sat})^{-1/3} - 1\right), \tag{35}$$

$$S_V^{sat} = \left(\frac{K_{FP}(1-Z_{vD}/Z_{iD})}{D_v^{rsd}Z_{vD}|B_{eD}|} - Z_{vD}\rho_D\right)^3 (4\pi N_V)^{-2}, \quad \tau_{sat} = \frac{S_V^{sat}}{Z_{iD}\rho_D D_v^{rsd}Z_{vD}|B_{eD}|} \tag{36}$$

where $N_V$ is the void number density, which has a strong effect on the swelling saturation parameters, as shown in the Fig. 9a. Typical $N_V$ values are different for different materials, which results in very different evolution of the void swelling, as shown in Fig. 9b for several typical cases. In pure Mo and Ni, high void densities are nucleated, and so swelling saturates quite rapidly as compared to stainless steels, which usually produce relatively low void number densities and are characterized by high swelling rates (up to about 1%/dpa). The evidence (for HFIR irradiation) for the saturation of swelling in Type 316 stainless steel [20], can be explained by the high concentration of helium in this case, which favors the void nucleation.

The saturation phenomenon is intrinsically connected with a void ordering observed in very different radiation environments ranging from metals [21-22] to ionic crystals [23], which attracts a lot of attention among physicists. The mechanisms of one-dimensional *interstitial transport* (currently, the most popular void ordering concept [2, 3, 23]) are still a subject of debate, especially in the case of ionic crystals. According to the present theory, ordering phenomena might be a natural consequence of the *energy transfer* along the close packed directions provided by focusons. This subject needs a special attention and will be considered in details elsewhere. Here we note that the maximum void lattice spacing along the closed packed directions in most cases is about of 100 *b*, which has been taken as the characteristic propagation range of a focuson in a perfect crystal throughout the calculations in the present work.

## 7. Discussion and conclusions

The main argument of the present work is that microstructural evolution and macroscopic deformation of materials under irradiation are essentially determined by the balance of *two main driving forces*, namely, Frenkel pair production in the bulk and Schottky defect emission from the microstructure components. And we have tried to demonstrate that the most salient radiation-induced effects cannot be understood well on the basis of only one of the two forces. The efficiency of the RSDE in the present model depends on only two new important parameters, namely, the characteristic propagation range of a focuson in a perfect crystal, $l_F$, and the kinetic factor, $k_F$, connecting the formation energy of "thermal" vacancies with that produced by a focusing collision. In the case of $k_F \to \infty$, or $l_F \to 0$ (Fig. 10a), the present theory is reduced to a "classical" rate theory, which regards ED primarily as the sinks for freely migrating point defects produced by irradiation in the bulk. The radiation-induced recovery and saturation phenomena described in the present paper cannot be understood within the classical framework. In real crystals, there is always present some concentration of impurities, which can act both as the focuson "sinks" that stop the focuson propagation, and "sources" that transform the more energetic but less long-ranged *crowdion* collision sequences into focusing collisions [24]. These opposite trends can be easily incorporated into the present theory resulting in rather ambiguous and non-monotonous impurity effects (Fig. 10b), which might be one of the major causes of the large difference in response to irradiation among the similar materials doped with different impurities.

Among the many other interesting implications of the present model, we would like to mention the *sub-threshold* electron irradiation effects, in which the electron beam energy is not sufficient for the bulk FP production but is well capable of producing s-defects. These effects have been observed for a long time, and in one of the first works on this subject [25], in 1964, Nelson was writing that under sub-threshold irradiation "point defects might conceivably be produced at dislocations or other extended defects, and almost certainly vacancies could be created at free surfaces of the foil" and that "care should be taken in interpreting experiments, such as the dissociation of defect clusters or climb of dislocations, solely in terms of the thermally activated release of point defects". We may conclude that now, almost 40 years later, we have followed this advice.

## Acknowledgements

This study has been supported by the NWO-NATO Grant #NB 67-292 and the AECL under the STCU Partner grant #P-049. The author is grateful to Drs Malcolm Griffiths, Anatole Turkin and Alexander Abyzov for many useful discussions.

Table 1. Material parameters used in calculations

| Parameter | Value |
| --- | --- |
| Atomic spacing, $b$, m | $3.23 \times 10^{-10}$ |
| Frenkel pair formation energy, $E_d$ | 30 eV |
| Crowdion range in a perfect crystal, $l_C$ | $10\,b$ |
| Focuson maximum starting energy, $E_F$ | 60 eV |
| Focuson range in a perfect crystal, $l_F$ | $100\,b$ |
| Focuson – dislocation cross-section radius, $r_{FD}$ | $5\,b$ |
| Thermal vacancy formation energy at a free surface, $E_{v0}$, | 1.8 eV |
| Thermal SIA formation energy at a free surface, $E_{i0}$, | 4.5 eV |
| Thermal PD formation energy at a dislocation with a stacking fault, $E_{sD}(SFE)$ | $E_{s0} + SFE\dfrac{\omega}{b}$ |
| Focuson-induced formation energy of PD at a dislocation with a stacking fault, $E_{sD}(SFE, k_F)$ | $E_{s0} + SFE\dfrac{\omega}{b}k_{FDs}$ |
| Dislocation kinetic factor for vacancies and SIAs, $k_{FDv}$, $k_{FDi}$ | $3 \times 10^2$, $3 \times 10^3$ |
| Vacancy formation energy at a GB perpendicular to the tensile stress, $E_{vG}(\sigma)$ | $E_{v0} - \sigma\omega$ |
| Vacancy formation energy at a dislocation with the Burgers vector parallel to the tensile stress, $E_{vD}(SFE, \sigma)$ | $E_{vD}(SFE) - \sigma\omega$ |
| Migration energy of SIAs, $E_{im}$ | 0.15 eV |
| Migration energy of vacancies, $E_{vm}$ | 1.3 eV |
| Migration kinetic factor, $f_c^m$ | 500 |
| Bulk recombination rate constant, $\beta_r$, m$^{-2}$ | $8 \times 10^{20}$ |
| Matrix shear modulus, $\mu$, GPa | 35 |
| Dislocation threshold glide stress, $\sigma_P$ | $10^{-3}\mu$ |
| Poisson ratio, $\nu$ | 0.33 |
| Atomic volume of the host lattice, $\omega$, m$^{-3}$ | $2.36 \times 10^{-29}$ |
| Relaxation volumes of vacancies and SIAs, $\Omega_v$, $\Omega_i$ | $-0.6\,\omega$, $1.35\,\omega$ |

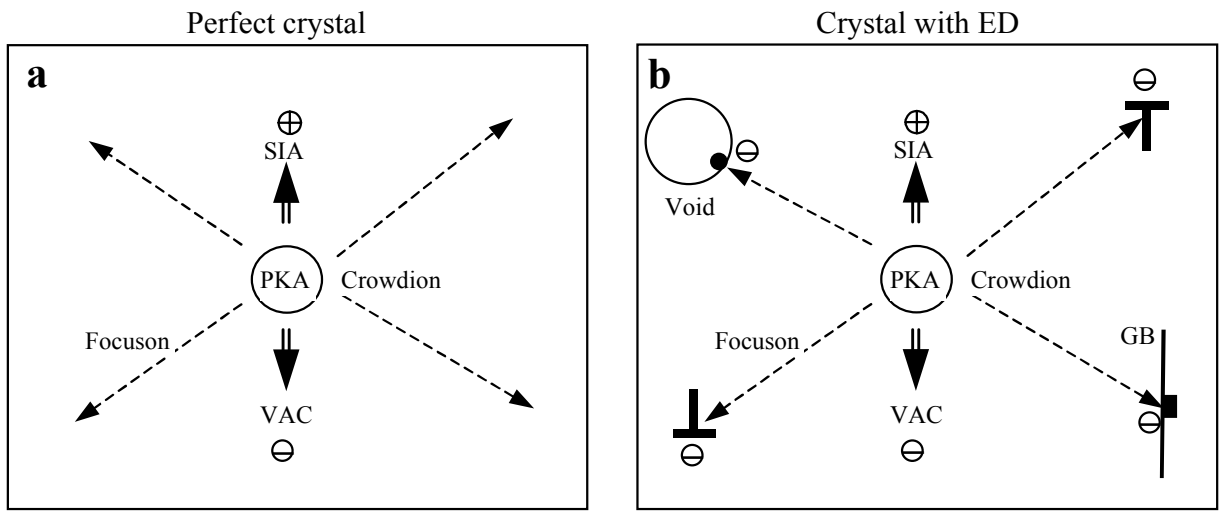

Figure 1. Illustration of different PD production schemes in perfect and real crystals: (a) Frenkel pair formation in the bulk by short-ranged crowdions; (b) vacancy formation at extended defects by long-ranged focusons

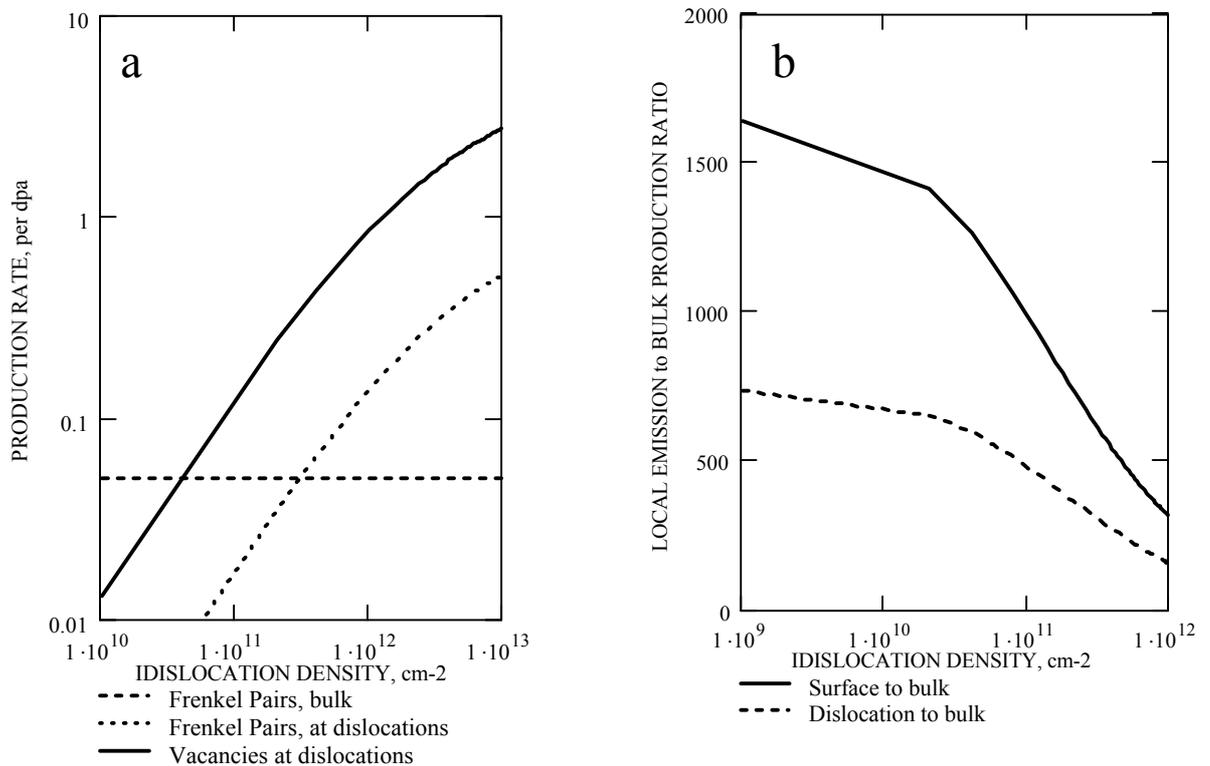

Figure 2. Comparison between production rates of FP and vacancies at dislocations per unite volume (a) and local (b).

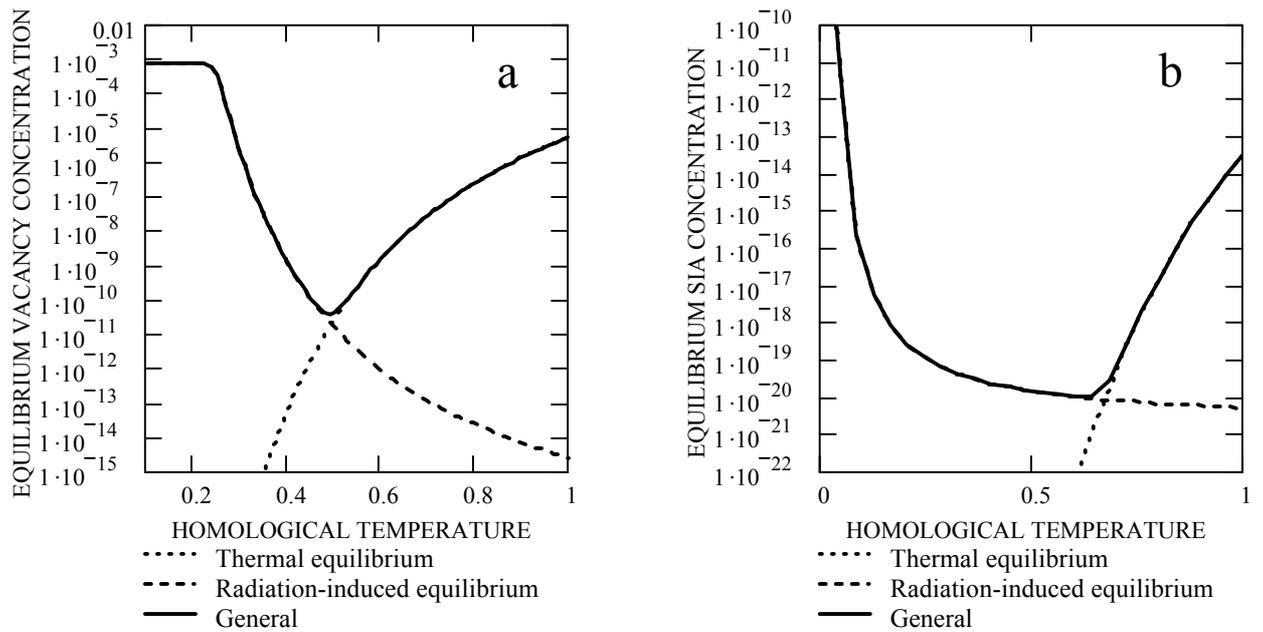

Figure 3. Comparison of thermal and radiation-induced equilibrium PD concentrations at dislocations for a typical neutron flux, $K = 10^{-6} dpa/s$

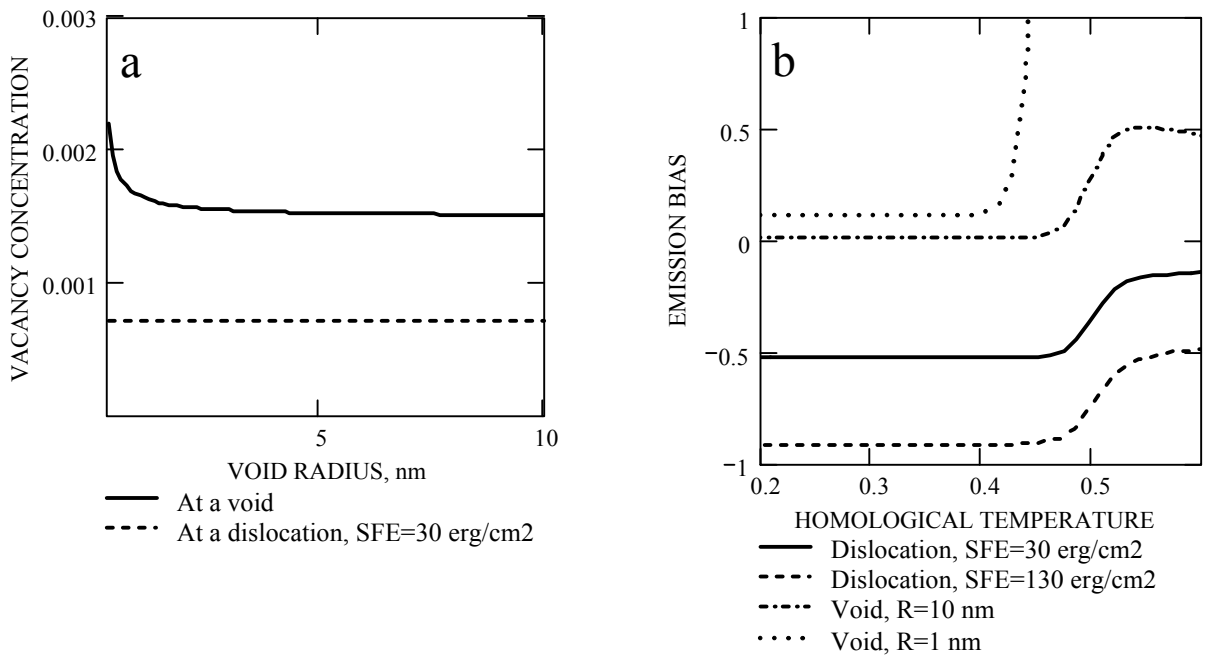

Figure 4. Vacancy equilibrium concentrations at voids and dislocations at $T < 0.5 T_m$ (a) and emission biases of voids and dislocations vs. temperature (b).

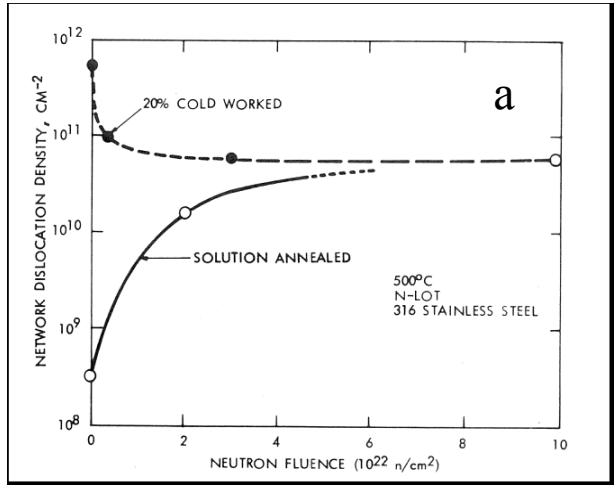 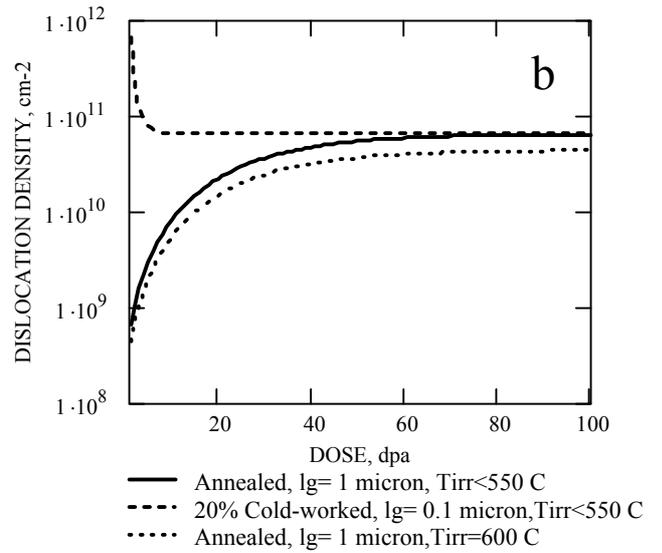

Figure 5. Comparison of experimental data on the dislocation evolution [15] (a) with the present theory, eq. (30), (b).

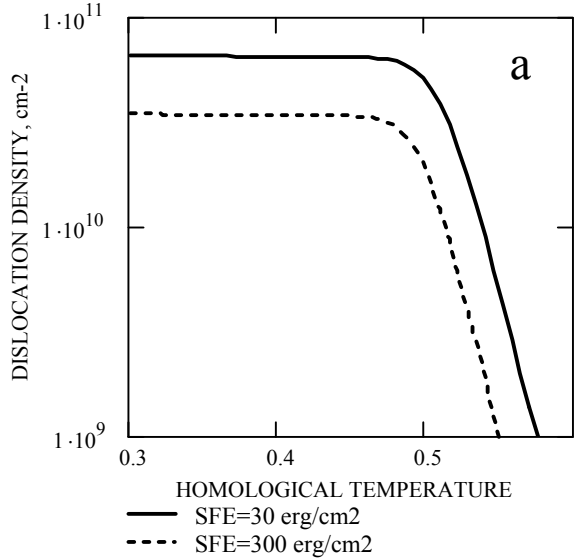 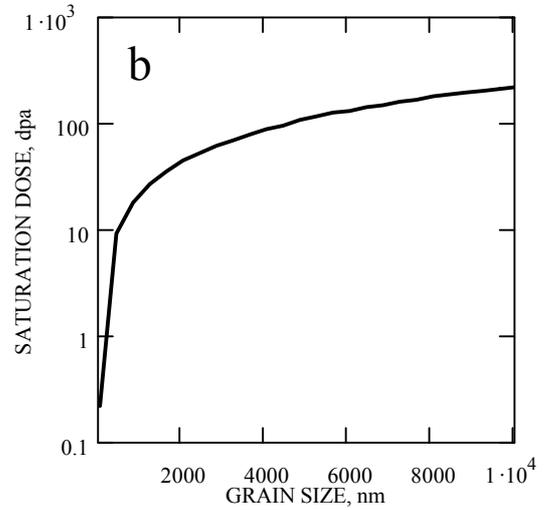

Figure 6. Saturation parameters of dislocation density vs. temperature and grain size given by the present theory, eq. (31).

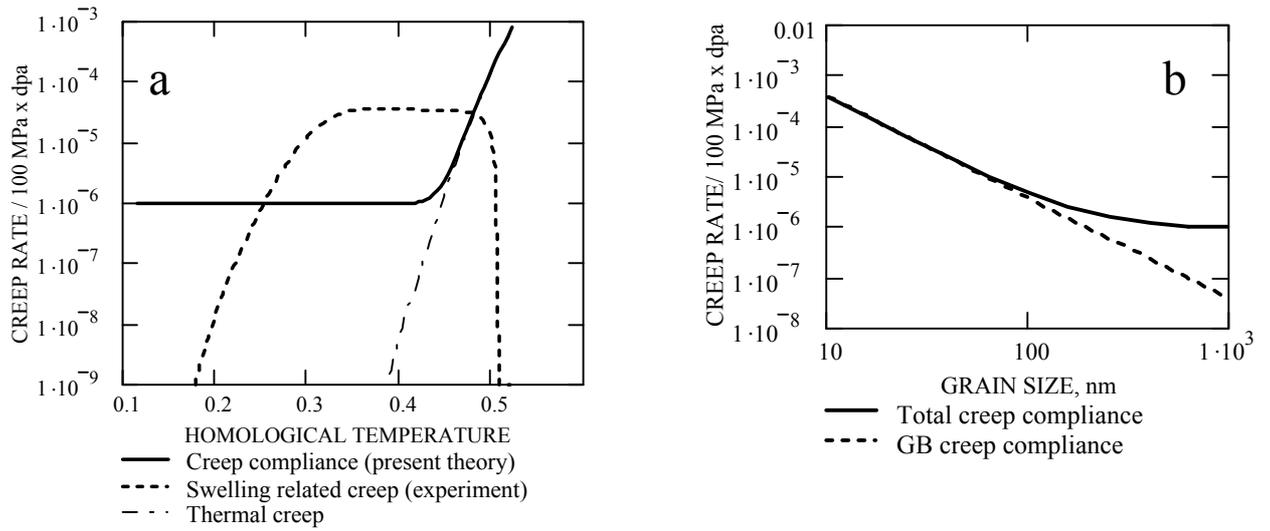

Figure 7. Creep rate dependence on temperature and grain size, predicted by the present theory, eqs. (33)-(34).

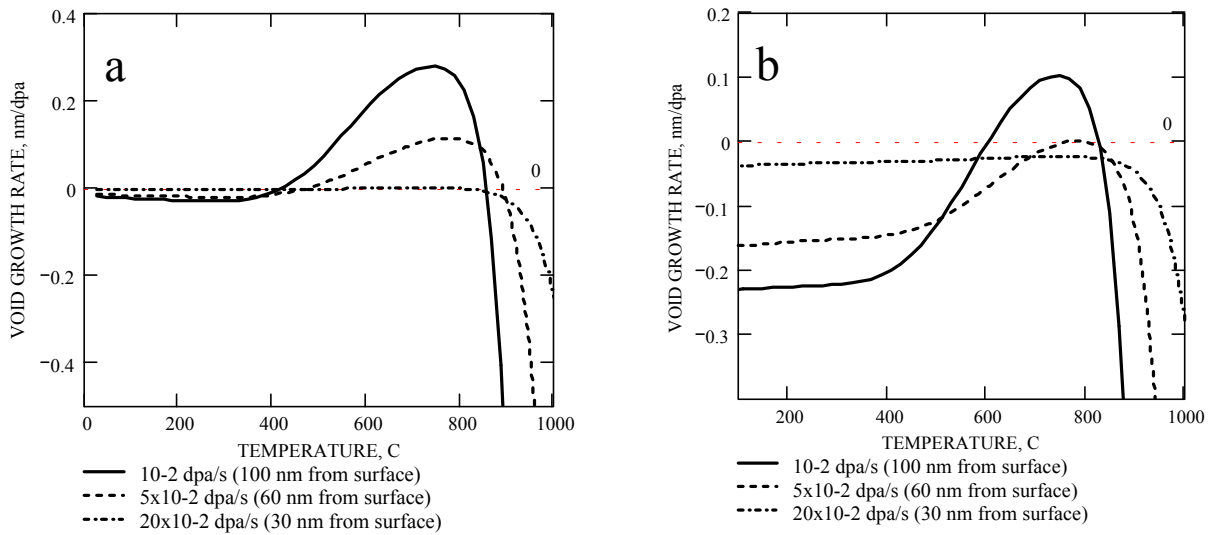

Figure 8. Void growth rates during $Ni^+$ ion bombardment of Ni according to

(a) conventional rate theory with account of production bias introduced by injected SIAs

(b) rate theory with account of RSDE

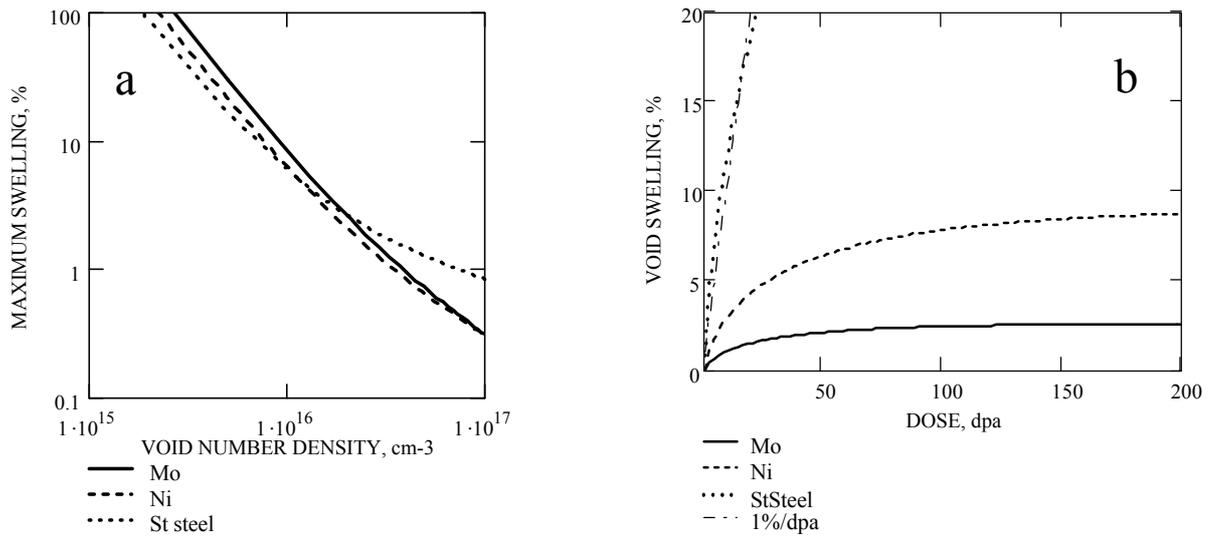

Figure 9. Saturation parameters of void swelling: (a) dependence $S_{sat}$ on void number density for materials with different SFE = 300 erg/cm2 (Mo); 130 erg/cm2 (Ni); 30 erg/cm2 (St steel); (b) Dose dependence of swelling for Mo ($N_V = 2\times10^{16} cm^{-3}$), Ni ($N_V = 8\times10^{15} cm^{-3}$) and St steel ($N_V = 2\times10^{15} cm^{-3}$)

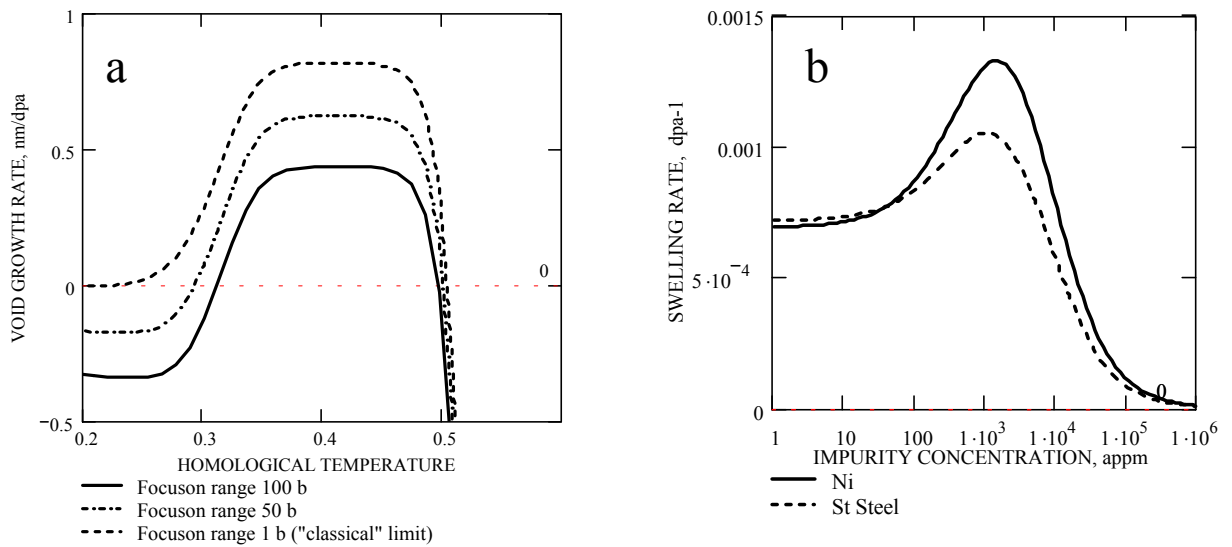

Figure 10. Swelling rate vs. the focuson propagation range in a pure crystal for different $l_F$ (a); in a crystal with impurities for $l_F = 100b$ (b)